# Building-in-Briefcase (BiB)

Kevin Weekly, Ming Jin, Han Zou, Christopher Hsu, Alexandre Bayen, Costas Spanos
University of California, Berkeley, Berkeley, CA 94720
{kweekly, jinming, hanzou, chriswhsu, bayen, spanos}@berkeley.edu

## ABSTRACT
A building's environment has profound influence on occupant comfort and health. Continuous monitoring of building occupancy and environment is essential to fault detection, intelligent control, and building commissioning. Though many solutions for environmental measuring based on wireless sensor networks exist, they are not easily accessible to households and building owners who may lack time or technical expertise needed to set up a system and get quick and detailed overview of environmental conditions.

Building-in-Briefcase (BiB) is a portable sensor network platform that is *trivially easy to deploy* in any building environment. Once the sensors are distributed, the environmental data is collected and communicated to the BiB router via TCP/IP protocol and WiFi technology which then forwards the data to the central database securely over the internet through a 3G radio. The user, with minimal effort, can access the aggregated data and visualize the trends in real time on the BiB web portal. Paramount to the adoption and continued operation of an indoor sensing platform is battery lifetime. This design has achieved a multi-year lifespan by careful selection of components, an efficient binary communications protocol and data compression. Our BiB sensor is capable of collecting a rich set of environmental parameters, and is expandable to measure others, such as $CO_2$. This paper describes the power characteristics of BiB sensors and their occupancy estimation and activity recognition functionality. Our vision is large-scale deployment of BiB in thousands of buildings, which would provide ample research opportunities and opportunities to identify ways to improve the building environment and energy efficiency.

## Categories and Subject Descriptors
C.3 [**Special-Purpose and Application-Based Systems**]: Real-time and embedded systems

## 1. INTRODUCTION
Indoor environment monitoring and control plays an important role in the operation of a building. One main goal of buildings is to ensure the safety and comfort of the occupants. Recent focus has been on minimizing energy consumption without compromising these goals. Tailoring services such as Heating, Ventilation, and Air Conditioning (HVAC), lighting, and electrical power, has the potential to save a significant amount of energy consumed. HVAC and lighting respectively comprise 48% and 22% of the total energy use of buildings in the USA [1]. In addition to environmental awareness, occupant-aware control schemes have been shown to save between 10-15% [3], 8.3-28.3% [4], or even 42% [5], depending on factors such as outdoor climate and control strategy. Having a more detailed view of building environment and its occupants opens the door to more energy savings as well as building services which are tailored to specific purposes and target groups.

Even though there are many solutions to monitor building indoor environments, three factors are limiting the applications, namely *portability*, *accessibility*, and *scalability*. Portability refers to the easiness to carry and move from one place to another, which is essential to improve the working efficiency and enlarge the set of monitoring sites. Accessibility is concerned about the data being easily reached and understood, which requires user-friendly interface, effective visualizations, and minimal technical expertise. Based on the previous properties, scalability is key for a solution to make impact, and it often leads to discovery of trends and patterns in large scale.

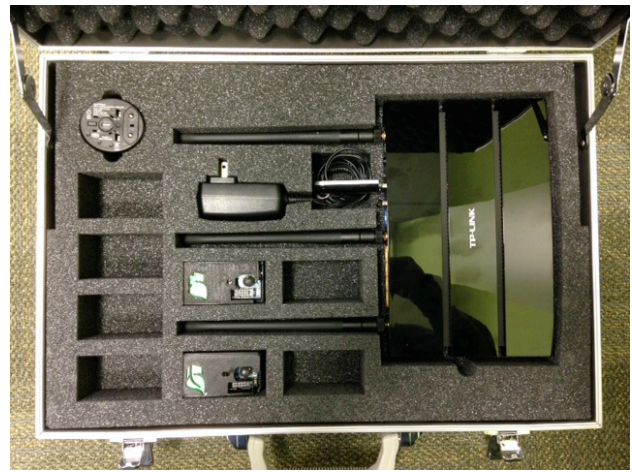

**Figure 1.** Building-in-Briefcase sensor, router, and briefcase. Each briefcase can hold eight (8) BiB sensors. By plugging in the power supply of the router, the sensors start collecting data which are forwarded to the internet data center.

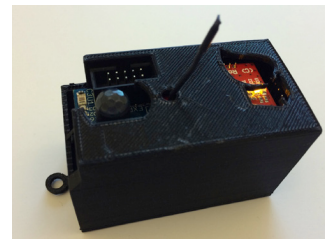

**Figure 2.** Building-in-Briefcase sensor up close. The cases are 3D printed to ensure proper functions of the sensors and easiness to attach to the interior surface.

Building-in-Briefcase (BiB), as shown in Fig. 1 and 2, is a portable sensor platform that supports scalable, continuous and long-term estimation of building environments and occupancy. The BiB sensor is a low-cost, battery-powered sensor which is small and light enough to be unobtrusively installed in an office space. It can optionally be outfitted with $CO_2$ or particulate matter sensors, which can be employed for the estimation of occupancy level [23]. The BiB sensor establishes reliable and high volume communications with the BiB router by a WiFi-compatible radio transceiver which supports data rates over 200 times higher than IEEE 802.15.4 technology. Additionally, we enabled 3G modem communication on the BiB Router for portability and adaptability to different building environments. The central database server is a single server designed to receive and efficiently store the time-series data collected by all of our BiB sensor networks. We also provide easy to access end-user customizable graphing and raw data access with a robust security model to address concerns of data privacy.

It is, therefore, the objective of this paper to describe the design and implementation of the Building in Briefcase. The rest of the paper is organized as follows. In Section 2 we present the system architecture. In addition to the BiB sensor described in section 4, there are three important aspects of the system; the wireless router, the database server, and data visualization. In Section 3 we provide a description of the communication protocol between the sensor and router. The hardware design is described in Section 4, including the microcontroller, radio module, power supply, sensing capability, and extension capability. The firmware design is presented in Section 5. In section 6, we evaluate the power efficiency of the BiB sensor and illustrate the application with examples from previous research. We conclude with some discussion and proposal for future work in Section 7.

## 2. SYSTEM ARCHITECTURE

There are three main components involved in measuring, communicating, and storing the environmental readings: the BiB sensor devices, the local wireless router, and the central database server. Fig. 3 illustrates how these agents are connected, as well as the relevant internal components of the local wireless router. The BiB sensors, which will be described in greater depth in section 4, connect to the wireless router using standard wireless 802.11 b/g secured with WPA2.

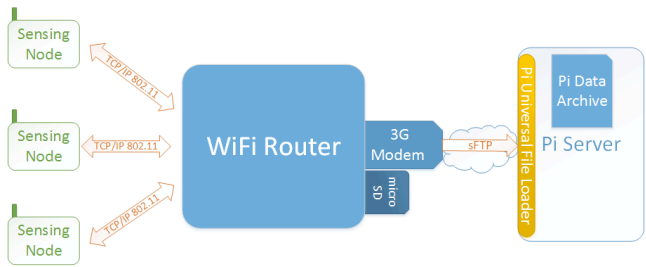

**Figure 3. System Architecture Overview.** The BiB sensors communicate with the WiFi router through TCP/IP 802.11 protocol. The data is forwarded to the PI Server by the 3G Modem which allows operation even without internet access infrastructure.

## 2.1 Wireless Router

The purpose of developing a self-sufficient local server is to support the idea of a "building in briefcase" sensor suite, where a sensor network can be rapidly deployed to initiate long-term monitoring, without the need for external infrastructure. Our latest solution is an off the shelf TP-Link wireless router running a custom configured OpenWrt build and custom Python processes. This combination of off the shelf hardware, customized OS build and custom code facilitates the following:

- Low power consumption and voltage of the router allows for short-term operation on portable lithium-ion batteries or longer term on solar power.

- Small size and no moving parts allow for the most flexibility in placement to optimize range and minimize obtrusiveness.

- Push-button simplicity. While a web-based configuration UI exists, a simple on/switch is the only interface we expect an end user to interact with.

- Extensibility. The availability of two built in USB ports allow for the addition of large quantities of local storage and choices for WAN connectivity.

### 2.1.1 OpenWrt and ROOter

OpenWrt is based upon the Linux kernel and primary used on embedded devices for routing network traffic. There are numerous optional software packages that can be added to the base OpenWrt distribution and it is straightforward to add additional components as needed.

The modifications to enable 3G modem communication were carried out as described by the ROOter community website [6].

Additional configuration was made to enable a USB flash drive to be supported and mounted as a Linux file system [7].

Since we planned to use Python as the primary language for custom code on the router we configured our OpenWrt build to include the full python 2.7 language environment. For secure-ftp (sFTP) communication we added python-paramiko and the dependencies python-crypto and python-ecdsa. Paramiko is a module that implements the SSH2 protocol for encrypted and authenticated communications with remote machines [8].

### 2.1.2 Python Socket Listener

The first custom Python process acts as the listener and socket handler for incoming connections from the BiB sensors. Upon the establishment of a client socket connection, data is read from the socket and decoded based upon our custom recordstore protocol described later. The decoded data is immediately written to a flat file on the USB Flash drive. Because we have our sensors configured to buffer data in memory and transmit only every 10 minutes, upon receipt of this accumulation of data the process writes this single connection to a file.

To prevent a single directory from getting bogged down with an unmanageable number of files, a directory is dynamically created for each day's worth of files.

### 2.1.3 Python File Distributor

The second custom Python process polls the specified directories on the USB flash drive for unsent files. Any unsent files are batched for multiple sequential puts across an sFTP connection made to the Database Server. Upon the successful sending of a file to the Database server, the file is renamed to indicate its sent status. To maintain space, but also provide for some data backup, after a configurable number of days has passed (currently 10 days) a day's specific directory and contained files are removed.

This current file transfer process has no trouble handling 10+ BiB sensors sending data on a 10 minute interval. However, due to the sFTP overhead incurred for each file sent, increasing the reporting interval to a much more frequent 15 seconds pushes the limits of the number of files we can send sequentially across a 3G modem.

If the number of BiB sensors associated to a single router grows significantly or if a more frequent reporting interval frequency is required, additional work can be done to aggregate and compress the component flat files prior to sFTP transfer.

## 2.2 Database Server

The central database server is a single server designed to receive and store the data collected by all of our BiB sensor networks. Maintaining all the collected data in a single repository allows for simplified data management and easy comparative analysis of collected data across all our deployment sites. To fulfill this need we needed database server and software that provides:

- High availability and solid stability. As a central component in the entire system, downtime or instability of this component would affect all sites at which a BiB sensor network was deployed and could not be tolerated.

- Scalability. While initial experiments involved two BiB networks with 6-10 sensors each, if the solution is deployed to hundreds or thousands of sites, we did not want to re-architect the backend solution.

- Efficient handling of time series data. With potentially thousands of sensors collecting data every second we need a storage solution that can intelligently store meaningful data while discarding noisy or redundant data.

### 2.2.1 sFTP Server

In order to get the data onto the Database Server we are currently running the SolarWinds free sFTP server [9], which receives the incoming sFTP connections from the BiB Routers and handles authentication and encryption. The sFTP process simply outputs the sent files into a specified directory.

### 2.2.2 PI System

The PI System is a collection of commercial Windows based software components that have been used in process and manufacturing industries for over 30 years [10].

The PI System is highly optimized to handle large quantities of time-series data while minimizing disk usage. It does this through configurable filters that allow noise below a given reading's typical margin of error as well as compression based upon a swinging door algorithm [11]. This flexibility allows us to maintain high sample rates without requiring unmanageable amounts of physical storage.

The PI System consists of a proprietary file-based database called the PI Data Archive that stores basic metadata and compressed time series data [12]. The PI Asset Framework is an overlay of additional metadata on top of data elements in the PI Data Archive and is built upon Microsoft SQL Server [13]. The Asset Framework allows users to build multiple hierarchical organizations of data into representations best suited for a given application. For example one view of a set of sensors may be broken out geographically while another view may be based upon building type (residential / commercial).

Data can be loaded into the PI Data Archive through any number of OSIsoft provided interfaces. In our particular case, to deal with a flat file we used the Universal File Loader (UFL) interface [14] that provides robust capabilities for handling flat files including files with varying message types per line. For example, the raw data output from the BiB sensors consists of some records containing occupancy or orientation change events while others contain the 7 time sampled data streams. The UFL interface handles these varying line types with simple configuration and no custom coding.

## 2.3 DATA VISUALIZATION

Of course the whole aim of the Building in a Briefcase project is to be able to view and analyze collected data to gain insight into building system and occupant behavior. Our goals for the basic data access were simple: provide easy to access end-user customizable graphing and raw data access with a robust security model to address concerns of data privacy.

### 2.3.1 PI Coresight

PI Coresight is a web based data visualization tool built by OSIsoft that is tightly integrated into the PI System Data Archive and Asset Framework [15]. It allows end users to easily generate plots of time series data for whatever time period or whatever data elements are of interest and allows dynamic zooming or panning of data across time. Also critical to our data visualization portal was the need to be able to control security at a granular level so that users of a BiB network in one building would only have access to their own data and not that of a BiB network deployed in a different location. PI Coresight and the PI System allows robust authentication options and role based data security to restrict or allow data access as the situation demands.

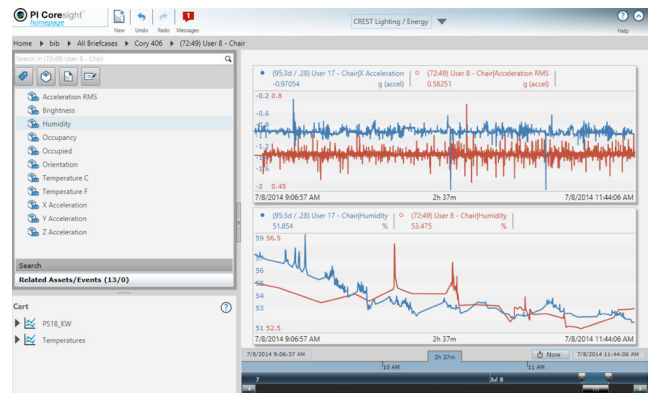

**Figure 4. Visualization Portal.** The left panel is organized in a structure from briefcase, to individual sensors, and to sensor measurements. User can only access the information with the given privilege. The right panel is a visualization of time series measurements, which can be arbitrarily zoom-in/out and overlay a list of measurements to compare the trends. The user simply needs to drag the corresponding sensor measurements in the left panel into the right to view the data in real time.

### 2.3.2 PI Excel Data-Link

For access to raw time series data, OSIsoft provides an add-in to Excel which allows for retrieval of data directly into an excel worksheet for further analysis [16]. Data can be retrieved in its compressed format where only meaningful changes to data values are reported. The amount of data returned when requesting compressed data is directly affected by the exception and compression settings set up in the PI Interface. Alternatively, data can be retrieved for time range with a value provided at each multiple of the requested time interval. In this case, if needed, PI does linear interpolation of the data to generate a value for every point in time.

Future work is planned to enable visualization of BiB sensor data overlaid onto floor plans or 3 dimensional building models to allow rapid comprehension of large amounts of interrelated data.

## 3. COMMUNICATION PROTOCOL

The communication protocol between the BiB Sensor and Wireless Router was custom built with several design goals. The protocol should be computationally simple allowing a microcontroller to process with minimal power. The protocol needs to be extensible but not overly complicated. Also it should support batching multiple readings into a larger packets sent at a frequency less than the sample frequency.

### 3.1 Lower Layer Considerations

The goal was not however to write an entire network protocol stack from the ground up so we take advantage of TCP/IP and its guarantees of reliable transmission. The TCP/IP stack extensively uses acknowledgement packets to validate the successful transmission of data, which is important to ensure that our sensor data is not lost, since we can later re-send the data if a transmission failed.

While TCP is chosen for its built-in mechanisms to ensure that data reliably reaches the server, this comes with a major drawback. Since TCP relies on many separate control packets to establish and tear down a connection, this requires that the radio to be active for the entire time. Thus, approximately one third of the sensor's battery life is expended by the radio alone, even at a very infrequent reporting rate. Future work will be to use the UDP protocol for which packet reliability is not enforced by the protocol. We can add a custom acknowledgment method that requires fewer overheads than TCP.

### 3.2 Message Framing and Description

Since TCP is a stream-oriented protocol, it guarantees that bytes are delivered reliably and in order (otherwise delivering an error). However, TCP does not provide framing utilities to designate discrete sets of bytes that should be treated as one message. The framing method chosen for our messages, utilizes a delimiter (we use the hex value 0x0A, i.e. the linefeed character) to indicate the boundaries between messages, by appending this delimiter to the end of every message. Thus, there is only one byte of overhead per message, and communications errors are generally recoverable for multiple message streams, as the next correctly received delimiter will be correctly interpreted.

### 3.3 Recordstore Data Format

Our sensor device relies on an inexpensive, and low-power microcontroller with limited computational resources. Since we must store sampled measurements to be later reported to the server, limited memory space can restrict the amount of power we can save by limiting the number of measurements that can be stored. For example, the ATmega1284 microcontroller that we use contains 16384 bytes of RAM, 2416 of which are used by our program. The average size of a message containing a measurement is 29 bytes, therefore, without compression, the maximum number of messages stored is 481. If a message is generated every 10 seconds, this corresponds to about 80 minutes of data stored. Thus, we developed *the recordstore data format* that the device uses to store messages to be later transmitted to the server. This provides simple and efficient compression, taking advantage of the fact that sensor data messages are very often the same length and have many repeated bytes, especially at high enough sample rates where environmental conditions do not change much between samples.

In addition to the messages being stored in the devices' RAM in this manner, messages are also transmitted to the server in this format (in fact the contents of the memory are simply "dumped" onto the communications channel). This therefore reduces the communication load. An additional advantage of the recordstore format is that it is stream-based, meaning that previously inserted bytes do not need to be changed (although they are referenced). The primary disadvantage of this format is that it will perform very poorly for rapidly changing data, such as measurements taken with very long sample intervals. In these situations, more overhead bytes may be added than the number that is saved by compression. Another minor disadvantage is that the compression technique limits the size of each message to 128 bytes, however, we currently have no need to send messages this long. In an experiment where 500 measurements were taken, once every 2 seconds, the recordstore method achieved a compression ratio of 1.5:1, or a reduction in memory requirements of 30%. Primarily, this serves to allow more samples to be stored in the device's memory prior to a report being sent. If latency is not a concern, this compression can significantly increase battery life due to fewer reports needing to be sent to send the same amount of data. Compression of time series data is an interesting and useful direction of study, for which the algorithm described is only an initial result. Future work will attempt to compress the data further since it is so important to the battery performance of the sensor.

Fig. 5 is an illustration of the data structure of the recordstore format. The memory block contains a list of records, each of which encapsulates one message. There are two types of records: template records and delta records, represented in Fig. 5 by the solid and hatched blocks, respectively. Additionally, to improve efficiency, a list of template pointers keeps track of the template records for quick referencing.

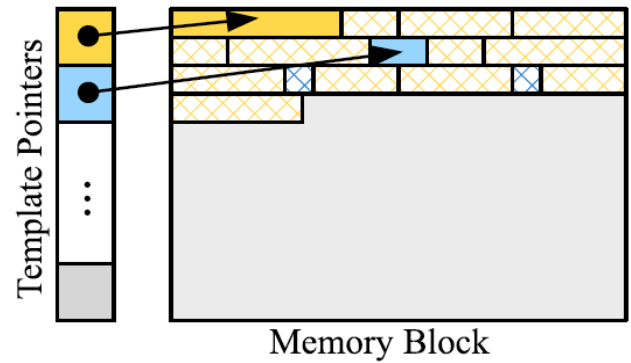

**Figure 5.** Diagram of the recordstore data structure in a typical scenario. There are two sets of records in this scenario, designated by the yellow and blue blocks. The solid yellow and blue blocks are the templates for that type of record and the hatched blocks are delta records which refer to the respective templates.

The exact format of the messages and the techniques used for encoding and decoding the messages are beyond the scope of this document but can referenced in the original thesis [2].

## 4. HARDWARE DESIGN

The objective of the hardware design for BiB is to design a permanent, reliable and easily deployable sensing infrastructure that can operate for many years inside a smart building. Since the design is from the ground-up, we have a great amount of flexibility in choosing state-of-the-art components available. The hardware design and the block-level diagram of the BiB sensor are demonstrated in Fig. 6 and 7 respectively.

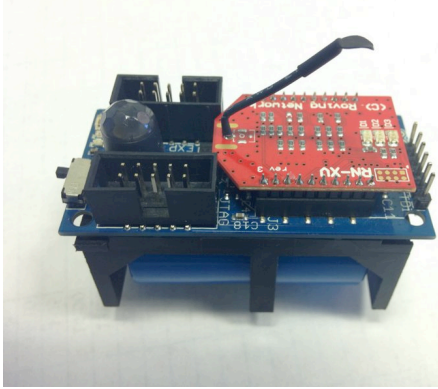

**Figure 6.** The hardware design of the BiB sensor. The RN-XV WiFi module is installed on top of the PCB board.

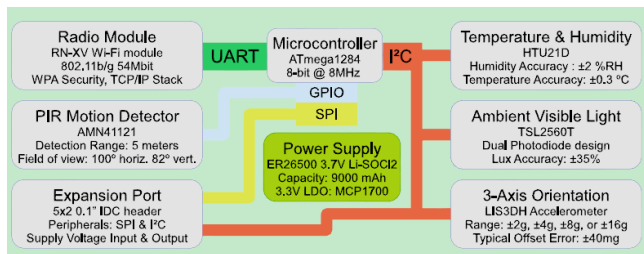

**Figure 7.** The BiB Sensor system diagram illustrating the major on-board instruments and expansion capabilities. Communications buses are indicted by solid-color paths.

### 4.1 Microcontroller

The logical center of the hardware design is Atmel's ATmega1284 microcontroller. This particular choice has a high amount of program memory (128KB) and RAM (16KB), which allows us to add in many code features without running out of space. More importantly, the microcontroller has the peripheral features required (I2C, SPI, and two UART ports), and the capability of low-power sleep while operating a 32.768KHz crystal. Even though there are many microcontroller families (e.g. TI's MSP430, Microchip's PIC) which satisfy our requirements, we chose the ATmega family due to its reliance on an open-source and free toolchain (GNU GCC) and extensive community support through the Arduino community (e.g. the community provides driver code for many sensors).

### 4.2 Radio Module

The radio module is another critical component for a wireless sensor. The function of it is to communicate with an endless variety of protocol and protocol combinations, considering the layered model of communications. We included a 20-pin socket, popularized by the ubiquitous XBee module, which is the connector footprint that is used by many of Digi's OEM XBee modules initially, because we did not seek to tie our design to a specific wireless module. Numbers of companies produce the XBee modules which adhere to the same connector footprint and signal locations due to its popularity. This allows us to evaluate these alternative offerings without a hardware redesign of our board. Eventually, as shown in Fig. 7, we chose Microchip's RN-XV [18] module as it proved to provide the most reliable communications in real deployments, owing to the retry-enforced reliability of the TCP protocol. Although the power consumption of this WiFi transceiver is up to 10 times greater than Bluetooth or IEEE 802.15.4 products, the high data rate (over 200 times faster) allows us to transmit many measurements in one "burst", thus limiting the time that the radio must be turned on.

### 4.3 Power Supply

The BiB sensor is powered via a 3.7V Lithium-Thionyl Chloride battery with a nominal battery life of 9Ah in the standard configuration. This particular chemistry is intended for powering long-lifetime devices, having a low self-discharge of less than 1% per year. Nevertheless, the capacity of the battery is reduced in some situations. One scenario is when pulling more than 2 mA of continuous current or more than 400 mA of pulsed current. The capacity of the battery is also reduced at lower current levels, such that at a 100 µA current draw (roughly the average draw of the sensor), the capacity is roughly 8Ah. A 100 µF capacitor is used in our design to lightly buffer the current draw during the times that the radio is active in order to overcome the problems associated with current.

Since the designed system voltage of BiB is 3.3V, therefore a linear regulator is leveraged to drop the 3.7V battery voltage to the required voltage level. The system can also be supplied with another power source of up to 6V and as low as 3.5V.

### 4.4 Sensing Capability

In this section we introduce the environmental variables that BiB is capable to measure. Table 1 below is a summary of the modules and performance parameters.

**Table 1** Summary of the sensing capability of the BiB sensor

| Environmental Parameter | Module | Performance |
|---|---|---|
| Temperature and Humidity | HTU21D | T accuracy: ±0.3℃ <br> H accuracy: ±2% RH |
| Ambient Visible Light | TSL2560T | Lux accuracy: ±35% |
| 3-Axis Orientation | LIS3DH Accelerometer | Range: ±2g, ±4g, ±8g |
| PIR Motion Detector | AMN41121 | Detection range: 5m <br> Field of view: 100° horiz., 82° vert. |

#### 4.4.1 Temperature and Humidity

Measuring temperature and humidity from the local environment is accomplished by Measurement Specialties' HTU21D [19] instrument. This instrument is attached to the shared I2C bus and can be queried to provide 12-bit and 14-bit digital readings of relative humidity and temperature, respectively. The stated accuracy of the humidity sensor is ±2%RH typical over the

20%RH to 80%RH range, and up to ±3%RH outside of this range. The maximum error tolerance is stated to be ±5%RH over the whole range. Although the instrument is calibrated at the factory, the relative humidity reading must be temperature-compensated to achieve the stated accuracy. The coefficients and formula needed for this correction are specified by the datasheet. The stated accuracy of the temperature sensor is typically $\pm 0.3$ °C and maximally ±0.4 °C over the range of roughly 5 °C to 60 °C which well covers the expected range of temperatures encountered indoors.

### 4.4.2 Ambient Light

We employed AMS' TSL2560 instrument [20] to measure the ambient visible light for BiB sensor. This design of the IC incorporates two light-sensing photodiodes: one measures visible and IR light from 300 nm to 1100 nm, and the other measures IR light from 500 nm to 1100 nm. Therefore, the reading from the second (IR only) photodiode can be used to compensate for the light energy that the first (visible and IR) photodiode measures, but is not visible to the human eye. An additional feature of the instrument is the ability to change the integration time of the on-board analog-to-digital converter (ADC), which can be changed depending on light conditions (e.g. a long integration time for dark environments). However, in our implementation, we have fixed the integration time to 101 ms. This instrument is attached to the shared I2C bus allowing the microcontroller to configure the instrument to read the 16-bit values the two photodiode measurements, which are used by the microcontroller to calculate a lux reading. We also used the interrupt feature of the instrument, which detects when the light level crosses above or below a preconfigured threshold. When enabled, this allows our sensor to timestamp events such as when the sensor is put inside a drawer or the lights are turned on and off.

### 4.4.3 Orientation

The orientation of the device is measured by the LIS3DH Accelerometer by ST Microelectronics [21]. It can also be utilized to measure high-acceleration "bumps", such as footsteps. Detecting orientation and bumps provides information in a mobile environment, such as if the device is in the pocket of a mobile human occupant. The number of footsteps taken and whether the person is sitting or standing can be extremely valuable pieces of information in an occupant tracking or occupancy estimation problem. In these situations, the accuracy of the instrument is not of large concern. For orientation purposes, the typical magnitude of acceleration we are measuring is 1 g, so the stated offset error of 40 mg represents about a 4% error. We use the interrupt capabilities of the device to inform the processor when the orientation has changed since the x, y, or z measurement axis is aligned with the gravity vector. The 3 16-bit values of acceleration are read from the device over the shared I2C bus.

### 4.4.4 Motion Detection

We measure motion with Panasonic's AMN41121 passive infrared (PIR) motion detector module [22]. This module utilizes a pyroelectric element to monitor small changes in infrared black-body radiation emitted by humans (7-14 μm).

The special property of this module is that there is a specialized lens which forms a pattern of discrete detection zones corresponding to one of four sensing regions of the pyroelectric element. Further circuitry within the device monitors changes the infrared measured by the four regions to determine whether a detection event has occurred. The sensor will detect when an infrared-emitting body eventually when a human moves across the detection zones, but remain insensitive to overall temperature increases or decreases within the field-of-view. The experimental results have shown that the sensor is also insensitive to an unmoving human. The field-of-view of the sensor is stated to be 100° along one axis and 82° along the other, and the detection range is at least 5 m from the sensor. Within this cone, there are 64 discrete detection zones distributed somewhat uniformly. A drawback of the module is that it needs at least 7 seconds for the circuit and sensor to stabilize before the detection is reliable. Therefore, we realistically cannot duty-cycle the sensor to save power, however, since the static power consumption is only 46 μA, we can leave the sensor powered and still achieve long battery lifetimes.

The module interfaces to the microcontroller via a single output which connects the attached signal to Vdd (HIGH) when a detection occurs. An external pull-down resistor pulls the signal to GND (LOW) otherwise. Although the datasheet does not specify the timing of the signal during detection, we have found that, the signal will stay HIGH as long as there is activity, but can sometimes intermittently go LOW, even while humans are moving in front of the sensor. Therefore, we require some intelligent interpretation of the signal, rather than simple assuming that a HIGH signal means humans are present and a LOW signal means humans are not present.

There are two calculated measurements to report: one is the occupancy percentage, which is the percentage of time (with a resolution of 1/256 seconds) that the signal was HIGH over the last sample interval. The second is the occupancy state changed event, which sends a value of 1 if the signal goes HIGH after being LOW for 10 seconds, or sends a value of 0 if the signal is LOW for 10 seconds after previously sending a 1. These two measurements are sufficient to determine whether the space in front of the sensor is occupied by one or more occupants.

## 4.5 Extension Capability

Besides the on-board sensors, BiB sensor also has the expansion port to allow virtually limitless expansion possibility to interface other sensors. The expansion port is a 10-pin male IDC connector with standard 0.1" spacing. The port exports an SPI Master interface, including clock (SCK), Master-Out-Slave-In (MOSI), Master-In-Slave-Out (MISO), and a single slave Chip Select (CS) line. These four signals can also be used as General-Purpose Input Output (GPIO) pins, including the ability to timestamp changes in voltage. The shared I2C bus is also available on the expansion port, composed of the Serial CLock (SCL) and Serial DAta (SDA) signals. It should be noted that all of these pins operate using 3.3V digital logic levels and are not intended to communicate with 5V logic levels. The expansion port includes four power-supply pins: two GND pins, VCC, which is the system voltage of 3.3V, and VBATT which is either 3.7V from the lithium primary battery, or can be used as a 3.5V to 6V supply input when the battery is not being used.

There are various external devices that have interfaced to the BiB sensor board to allow other types of measurements. One of the potential integrated external devices is $CO_2$ sensor. Sensing $CO_2$ has widely usage in intelligent building research, particularly in occupancy estimation [23]. We have successfully interfaced the K-30 $CO_2$ sensor from CO2Meter [24] to the expansion port of the BiB sensor board. The K-30 sensor achieves an accuracy of $\pm$ 30 ppm $\pm$ 1% using a self-calibration procedure called Automatic Baseline Correction (ABC) which adjusts the readings such that the lowest value in the last 7.5 days is equal to 400 ppm. An external power source is required since the K-30 sensor requires a power input of 4.5 V to 9 V at an average current of 40 mA and maximum current of 300 mA. We connect the VBATT input of the BiB sensor to the power supply of the K-30 device, so that they share the same power source.

## 5. FIRMWARE DESIGN

The firmware design of BiB system is mainly completed on the ATmega microcontroller. There is no underlying operating system for this design. The ATmega microcontroller was programmed in C code and debugged using the on-board JTAG connector. The code is organized logically as follow:

- Device drivers – code which knows how to configure and extract data from the instruments.

- Peripheral drivers – code which knows how to configure and use the ATmega hardware peripherals which are shared between multiple modules. Drivers for hardware peripherals which entirely used by one code module are usually included within that module.

- Radio drivers – code which knows how to configure and use the attached radio transceive.

- Utilities – generic, non sensor-related, utilities such as CRC generation and a task scheduler.

- Sensor Logic – Computational code, such as scheduling when samples are taken and reported.

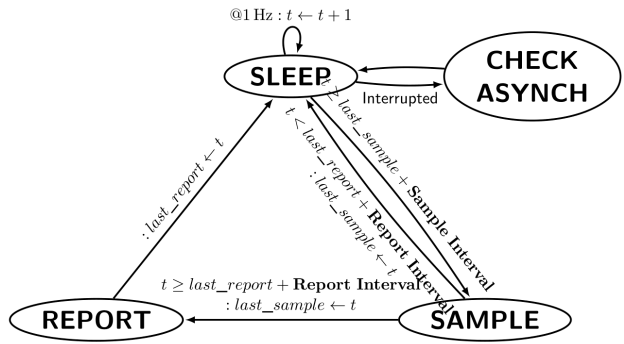

**Figure 8.** State machine diagram of the firmware running on the ATmega microcontroller of the BiB sensor. The variables $t$, *last_sample*, and *last_report* represent time variables and initialized to 0. **Sample Interval** and **Report Interval** are configuration variables.

The operation of the sensor can be summarised as a finite state machine with four states. As shown in Fig. 8, they are:

- The machine begins in the **SLEEP** state, which represents the lowest-power state.

- The machine may be interrupted (e.g. by the PIR sensor) and transition to the **CHECK ASYNCH** state which interprets the interrupt and possibly creates a Sensor **Data Report**.

- The machine may also transition to the **SAMPLE** state if it is time to read the next periodic sample and create a **Sensor Data Report**. A transition to this state triggers the sampling schedule to execute.

- Finally, the machine transitions to the **REPORT** state if it is time to transmit the stored samples to the server over the radio.

The processor must activate the peripheral instruments when the machine transitions to the **SAMPLE** state, instruct them to make a conversion, wait for the conversion to complete, then retrieve the conversion result. The processor can be simultaneously communicating with another instrument to make the system more efficient when it is waiting for a conversion to complete for one instrument. In addition, a simple real-time scheduler is provided for the device drivers to use to schedule their operation to make this process straightforward. This sampling scheduler is triggered when the **SAMPLE** state is entered. The scheduler allows drivers to specify the earliest time that a task should be executed.

**Table 2.** Standard Schedule of tasks executed to take one sample

| Time (ms) | Component | Description |
|---|---|---|
| 0 | Reporting | Start constructing new **Sensor Data Report** |
| 1 | Temp/Humid | Start humidity conversion |
| 1 | PIR | Calculate PIR occupancy percentage value and reset |
| 1 | Light | Wake light sensor to convert |
| 1 | Accelerometer | Read latest acceleration |
| 17 | Temp/Humid | Read converted humidity and start temperature conversion |
| 67 | Temp/Humid | Read converted temperature |
| 106 | Light | Read converted ambient light |
| LAST | Reporting | Store **Sensor Data Report** into record store memory |

The sampling schedule for instruments on the BiB sensor is demonstrated in Table 2. The first column gives the time offset from when the sample is triggered. As presented, the light and the temperature/humidity instruments require a conversion time which is enforced by the scheduler. When more than one task is specified to execute at the same time, such as at the 1 ms time offset, the scheduler executes them in the order they were placed into the queue. As shown in the last row of Table 1, there is a special value of time offset, LAST, which instructs the scheduler to execute the task after all other tasks. In order to save power, the scheduler turns off the CPU when no tasks need to be run.

## 6. APPLICATION AND EVALUATION

In this section, we perform evaluation of the power efficiency of the BiB sensor, which is an essential aspect for building monitoring. Several potential applications, including occupancy estimation and activity recognition, which rely on the BiB for experiments, are described, as a demonstration of the portability and accessibility of the BiB platform.

### 6.1 Power Efficiency

The power efficiency (battery lifetime) is one of the critical aspects when we design the BiB sensor to make it to achieve a multi-year battery lifetime to reduce the maintenance cost of the network. In order to bring down the average current consumption to a target of less than 200 µA, several strategies have been employed.

For instance, we select the components with low current requirements only. This helps us reduce the amount of energy consumed, as well as reduce the peak load demanded to be supplied by the linear regulator and battery. Linear regulators with higher peak current capability also typically have a higher leakage current. The high peak current draws can damage or reduce the effective capacity of the battery. Since the most vital component of the BiB sensor is the radio transceiver, we did the selection of it cautiously. Eventually, we selected the Microchip's RN-XV module instead of others such as the XBee-PRO and XBee Series 6 (WiFi), because of their high (over 300 mA) peak consumption.

Furthermore, we create power consumption worksheet to determine the feasibility of having a multi-year battery life. We list the power requirements for each device during their sleep and active modes, and the typical amount of time required to be active to perform their functions. After that, we simulate and plot (as shown in Fig. 9) the battery lifetime of the BiB sensor powered by the 3.7V Lithium-Thionyl Chloride battery, over varying values of sample intervals and report intervals. The current consumption from the light sensor, humidity and temperature sensor, accelerometer, PIR sensor, microcontroller, and radio is simulated. In addition, we also simulate the reduction of battery capacity at low average current draw.

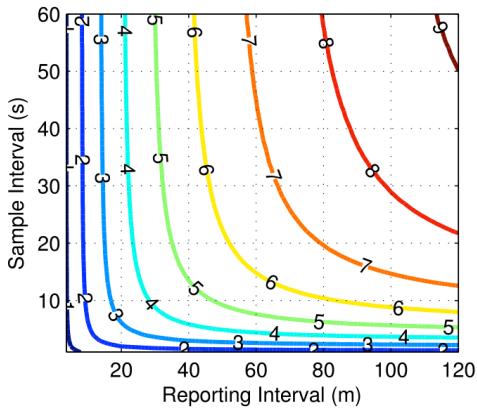

**Figure 9.** Battery Life (surface contours, in years) given by varying the amount of time between (y-axis) and time between transmitting the data to the server (x-axis).

Based on our analysis and evaluation, the BiB sensor can achieve a battery lifetime of over 5 years by using a 10 second sample interval and 60 second reporting interval. In this configuration, the average current consumption is 168 µA, and the effective battery capacity at this current is 8.03 Ah. The amount of current being used for communication is 56 µA, 57 µA for sensing, and 47 µA for processing. Another 8 µA is used for inactive devices while they are in their respective sleep modes

## 6.2 Example: Occupancy Estimation

Various methods have been employed for occupancy estimation, such as passive infrared (PIR) sensors, ultrasound sensors, and magnetic switches. These types of sensors provide accurate detections of occupants; however, the information they provide is limited. For instance, these light-based and ultrasound-based sensors usually have a small detection volume and cannot distinguish the number of occupants or the amount of activity that is occurring [25].

To explore techniques which do not have these limitations, the occupancy level of indoor spaces is directly estimated by measuring the $CO_2$ concentration [23] with BiB sensors. The dynamics of the $CO_2$ concentration in the room is modeled using a convection PDE with a source term which is the output of a first-order ODE system driven by an unknown input which models the human's emission rate of $CO_2$. The source term represents the effect of the humans on the $CO_2$ concentration in the room. In the experiments, a delay is observed in the response of the $CO_2$ concentration in the room to changes in the human's input. For this reason, the source term is a filtered version of the unknown input rather than the actual input. It is assumed that the unmeasured input from the humans has the form of a piecewise constant signal. This formulation is based on our experimental observation that humans contribute to the rate of change of the $CO_2$ concentration of the room with a filtered version of step-like changes in the rate of $CO_2$. Fig. 10 shows a typical trace of $CO_2$ concentration when the occupancy changes.

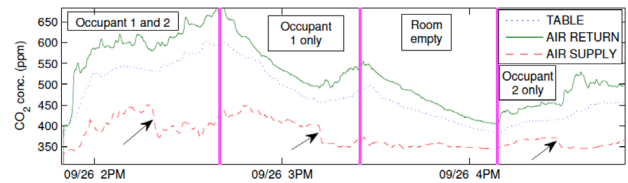

**Figure 10.** $CO_2$ concentrations over 3 hour experiment. Measurements from 3 locations in the conference room, namely the air supply and return vents, and the conference table at the center of the room, are shown in the plot. Magenta lines indicate when occupancy changes occurred. The arrows indicate the time instants at which the ventilation rate increases.

As can be seen, the extension capability of BiB sensors to measure $CO_2$ concentration is directly applied in this study. At the conclusion, a PDE-ODE model is developed that describes the dynamics of the $CO_2$ concentration in a conference room. An observer is designed and validated for the estimation of the unknown $CO_2$ input that is generated by occupants, which acts as an intuitive proxy for the number of occupants breathing in the local air space.

## 6.3 Example: Activity Recognition

Building intelligence encompasses its ability to sense and understand the activities of occupants to interact with them and achieve goals like comfort and energy efficiency. Individuals perform various activities inside the building. This information, when made available to the building automation and control system, can be very useful. For example, the PMV model proposed by Fanger and adopted by ASHRAE as the primary standard for thermal comfort takes occupant metabolic rate as the most important factor, but it has been widely regarded as the most difficult parameter to measure.

The BiB sensor was adapted into a watch to conduct wearable sensor studies [26]. The goal of using BiB was not to be smaller than the current offerings, rather to be small enough to enable these studies.

It was shown via experimentation that indoor occupancy activity can be recognized and classified by leveraging the environmental measurements, including temperature, humidity, and lighting level. Features including temperature gradients and standard deviation, humidity standard deviation, lighting levels are proposed for activity and location recognition. The features are statistically shown to have good separability and are also information-rich. Fusing environmental sensing together with acceleration is shown to achieve classification accuracy as high as 99.13%. For building applications, this study motivates a sensor fusion paradigm for learning individualized activity, location, and environmental preferences for energy management and user comfort. The capability of BiB to measure temperature, humidity, light level, and acceleration is demonstrated in this study.

Table 3. Feature table of samples of environmental sensors.

| | Digi XBee Sensors [71] | Telos Platform [66] | Powercast WSN-1101 [72] |
|---|---|---|---|
| **Design** | 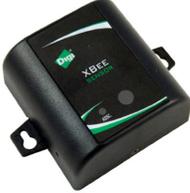 | 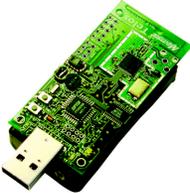 | 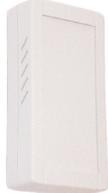 |
| **Measurements** | Temperature, humidity, ambient light | Temperature, humidity, ambient light | Temperature, humidity, ambient light, $CO_2$ (optional) |
| **Battery Type** | 3 Alkaline AA Cells (4.5V 2700mAh) | 2 Alkaline AA Cells (3V 2700mAh) | Integrated Lithium OR Radio Power Transfer |
| **Battery Lifetime** | 1.5 years (1/30 Hz rate), 2.5 years (1/60 Hz rate), 6 years (1/3600 Hz rate) | 3 years (1% duty cycle) | 25+ years (battery), perpetual (RF-powered) |
| **Communications** | 2.4 GHz IEEE 802.15.4 ZigBee mesh network | 2.4 GHz IEEE 802.15.4 TinyOS mesh network | 2.4 GHz IEEE 802.15.4 Proprietary mesh |
| **Cost** | 109 USD | 110 USD | 200-400 USD |

## 7. RELATED WORK

Prior to designing a custom hardware solution, we first evaluated alternative commercial solutions, as shown in Table 3. Some platforms, such as the Digi XBee Sensors [27], are commercially packaged, produced and sold through major distributors, which makes it extremely quick and convenient to build a platform upon. However, none fulfilled all our requirements of variables sensed, battery life, or cost. Moreover, proprietary solutions prevent us from extending the sensors to measure more variables, installing experimental networking protocols, and are also vulnerable to becoming unsupported by the company.

There are also many WSN-centric development kits intended for research and development, perhaps the most referenced and studied being the Telos and TelosB platforms [28]. These somewhat unfinished products and require some development of software, hardware, and mechanics to fit the parameters of our deployment. They also rely on an experimental and changing code base, such as TinyOS, and require a higher level of knowledge to configure, as opposed to simple and user-friendly interfaces such as the X-CTU utility by Digi. However, with this cost comes the benefit of being able to keep up with improving protocols and a high amount of customization.

Finally, there are proof-of-concept and short-run products such as the Powercast WSN-1101 [29] (and derivatives), which featured remarkable improvement in battery life and package size. Sensor nodes like these are usually developed by companies to demonstrate an underlying technology innovation (in this case, wireless charging, low-power micro-controller, and energy storage improvements). In many cases, the true product is an OEM module that is sold to other companies to integrate into a commercially packaged product.

The design of Building-in-Briefcase (BiB) differs from them by making the system trivially easy to deploy, enabling easy access to a variety of environmental parameters through effective visualization and data aggregation, which opens up many possibilities for large scale data mining for building managers and researchers alike.

## 8. CONCLUSION

A thorough understanding of indoor air quality and occupancy distribution is necessary to achieve the potential of improved building energy efficiency and reduced environmental impact, without compromising occupant comfort, safety, security and productivity. We should consider the grid, the building and its occupants, as an ecosystem in order to achieve the cooperative optimization of the interactions between them. The objective of the Building in Briefcase system is to provide a portable sensor platform for indoor environment monitoring and optimal building energy management that is simple to deploy in any building environment.

In this paper, innovative design and implementation of the BiB are demonstrated. The BiB sensor we have developed is small-size, low-cost, battery-powered and light enough to be unobtrusively installed in any type of indoor environment. It is capable of collecting a rich set of environmental parameters, and is expandable to measure others, such as $CO_2$. The hardware design of the sensor, including the microcontroller, radio module, power supply, sensing capability, and extension capability were selected to maximize the *portability* of the BiB system. The environmental data collected by BiB sensors are wirelessly sent to the BiB router via TCP/IP protocol and Wi-Fi technology. Then the data are securely forwarded to the central database through the 3G network. As we described in the paper, the communication protocol, the database server, the data visualization, the BiB router and the implementation of 3G modem communications contribute to the full-scale *accessibility* of the BiB system. These unique *portability* and *accessibility* features of BiB make it possible to be deployed in varied building environments. They also allows for scalability. The implementation and impact of the BiB system will be enormous and profound in the field of academic research, as well as in the practical fields of building management.

Our vision of BiB is to transform the practice of building environment and energy management through large-scale deployment. In the future, we will deploy a large quantity of BiB in buildings in Singapore to monitor the indoor air quality and environment. We will also keep adding services on top of the platform to improve its functionality and user experience.

## 9. ACKNOWLEDGMENTS

This research is funded by the Republic of Singapore's National Research Foundation through a grant to the Berkeley Education Alliance for Research in Singapore (BEARS) for the Singapore-Berkeley Building Efficiency and Sustainability in the Tropics (SinBerBEST) Program. BEARS has been established by the University of California, Berkeley as a center for intellectual excellence in research and education in Singapore.